\documentclass{sf2a-conf2026}
\usepackage{graphicx}
\usepackage{hyperref}
\usepackage[]{natbib}  
\usepackage{epstopdf}

\def\BibTeX{{\rm B\kern-.05em{\sc i\kern-.025em b}\kern-.08em
    T\kern-.1667em\lower.7ex\hbox{E}\kern-.125emX}}
\bibpunct{(}{)}{;}{a}{}{,}  %%%%%%%%%%%%%  A&A bibliography style
\hypersetup{colorlinks=true,linkcolor=blue,citecolor=blue,urlcolor=blue}

\newcommand{\teff}{\ensuremath{T_{\rm eff}}}
\newcommand{\fbol}{\ensuremath{f_{\rm bol}}}
\newcommand{\logg}{\ensuremath{\log g}}

\begin{document}

\TitreGlobal{SF2A 2026}

%%-----------------------------------------------------------------
%%      the top matter
%%

\title{Stellar physics at the core of modern astrophysics}

\runningtitle{Stellar physics at the core of modern astrophysics}
%Stars at the core of astrophysics

\author{O.\,L. Creevey}\address{Université Côte d'Azur, Observatoire de la Côte d'Azur, CNRS, Laboratoire Lagrange, Bd de l'Observatoire, CS 34229, 06304 Nice cedex 4, France}

%% Keep this line, even if the page will be settled afterwards.
\setcounter{page}{237}

%%-----------------------------------------------------------------

\maketitle

%%-----------------------------------------------------------------
%%        The abstract
%% 
%%  Warning!  within the abstract:
%%  - do not use macros. 
%%  - do not use commands like: \cite, \citet, \citep ... etc.

\begin{abstract}
Stellar physics provides an essential foundation for interpreting a wide range of astrophysical phenomena. In recent years, observational and theoretical advances have highlighted the central role of stellar physics at the interface with other fields, such as the study of exoplanets, galactic archaeology, stellar populations, and large-scale surveys like the Gaia and PLATO missions. In this article, I provide an overview of these interfaces, focusing on recent developments and examples of research conducted in France. The goal is to illustrate how modern stellar physics contributes to an integrated understanding of stars, planets, and the Milky Way, as well as to highlight some current challenges and future prospects.
\end{abstract}

%% Insert the keywords (to appear in the ADS indexing)
%% Keywords must be separated by a comma
\begin{keywords}
stars, exoplanets, Galaxy, fundamental properties, ages
\end{keywords}

%%-----------------------------------------------------------------

\section{Introduction}
\begin{figure}
 \centering
 \includegraphics[width=0.8\textwidth,clip]{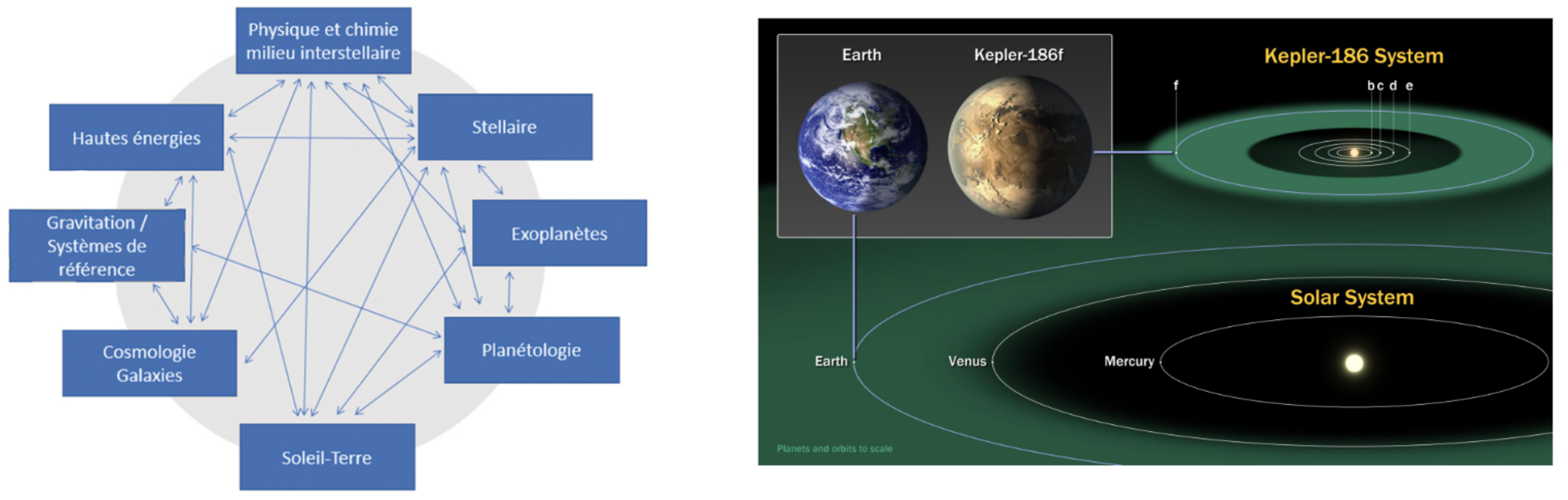}
%% Note the ABSENCE of the extension .pdf  !
  \caption{
  {\textbf{Left:}} Principal relations between the seven Action Themes of the National Programme of Astronomy and Astrophysics section and Planetology.  Credits: \citet{CoNRS2024Section17}.
  {\textbf{Right:}} An artist's view of the Kepler 186 exoplanetary system.  Credits:  NASA/Ames/SETI Institute/JPL-Caltech.
}
  \label{creevey:fig1}
\end{figure}

%%---------------------
Stellar physics is at the core of many astrophysical questions.  This is illustrated through the Conjecture Report of the CNRS Section 17 (now 19) in 2024 describing the interactions between the various National Program Action Themes in Astronomy and Astrophysics and Planetology, where each topic is tightly related to stellar physics, see Fig.~\ref{creevey:fig1} left panel.   
A figure in a national document, however, may not relate to many people so here I give a clear example.  Large spectroscopic surveys for Milky Way studies provide large homogenous samples of chemical abundances and metallicities which can ultimately be used to test diffusion processes in stars.  Or space-based missions, such as Gaia, provide kinematical information which enables us to identify binarity, groups, and populations including cluster memberships, and ultimately provide constraints on stellar fundamental properties. 
The input from Milky Way studies (Cosmology-Galaxy theme) clearly provides important constraints to help us gain a more deeper understanding of how stars work.  
In this work, however, I will focus on the other direction: how advancing in stellar physics has an important role to play at the core of other themes.

Fig.~\ref{creevey:fig1} right panel illustrates the Kepler-186 exoplanetary system from an artist's point of view compared to our own Solar system.    What is striking here is the comparison of the scale of that system compared to our own and it immediately leads one to realise that any knowledge we would like to gain on that exoplanetary system depends entirely on the knowledge of the planet host star; its mass which determines the speed of the planets in the orbits, its radius, which allows us to estimate the planetary radius, its luminosity, which dictates which planets may be habitable, and its age, which is a key ingredient for unravelling the exoplanetary evolutionary scenarios.  Recognising the importance of knowledge of the host, The ESA/PLATO mission \citep{2025ExA....59...26R} has at its core program the precise characterisation of stars.  Looking out towards our own Milky Way, the large scale surveys mentioned earlier can only provide chemical abundances because of our knowledge of stellar atmospheric models, and interpretation of stellar populations requires the knowledge of the stellar age and mass, which depend entirely on our understanding of how a star works.
Putting all of this together there two key concepts that emerge: deriving precise and accurate fundamental stellar properties (observational), and better understanding of stellar structure, evolution and atmospheric models (theoretical). % One the one side we have a more observational approach, while the other focusses on models. 
With these two aspects in mind, Sect.~\ref{creevey:sec:obs} presents a small overview of some recent advances in the observational aspects on fundamental parameters (radius, mass, age, ...), while Sect.~\ref{creevey:sec:model} describes some recent work on advances in modelling. 

\section{Focus on advances in observations}\label{creevey:sec:obs}
%%-------------------------
In this section, I elaborate on some of the recent advances in observational aspects on fundamental parameters.  

\subsection{Benchmark stars}\label{creevey:sec:benchmark}%%-------------------------
A benchmark star is a star for which its fundamental properties can be measured with little intervention of models, ideally none, and that when measured repeatedly or in different ways we obtain the same result. 
%By fundamental properties we mean the intrinsic properties of the star; its radius, mass, luminosity, effective temperature \teff, rotation, chemical composition.  
It also implies that we know one or more of its fundamental properties accurately and not just precisely.  The interest in these stars is that by constraining the 
properties observationally, we can then use them to test different theoretical scenarios in order to improve our understanding of the physical and chemical processes taking place in the interior or atmospheres of stars.
These stars can additionally be used as a reference set for testing pipelines designed to work on large datasets, and/or for calibrating or validation of the output of the pipelines.
Optical interferometry plays a particularly important role here because it allows one to measure angular diameters or precise orbits of binaries.   These measurements coupled with complementary measurements such as a parallax, enable a model-independent determination of fundamental properties, for example the radius or its \teff.   
%The French community are particulary well-known world-wide for their expertise in optical interferometry. 
One well known example, apart from the Sun, is the bright metal-poor giant {HD\,122563}.  Its \teff\ has been measured using interferometry, spectroscopy, and the infra-red flux method and all methods agree to 40 K \citep{2012A&A...545A..17C,2014MNRAS.439.2060C,2018MNRAS.475L..81K,2024A&A...682A.145S} With such tight constraints, non-LTE line formation in stars or convective processes can be tested  \citep{2016MNRAS.463.1518A,2018MNRAS.475.3369C,2024A&A...682A.145S,2024A&A...689A.243C}.

\begin{figure}
 \centering
 \includegraphics[width=0.45\textwidth,clip]{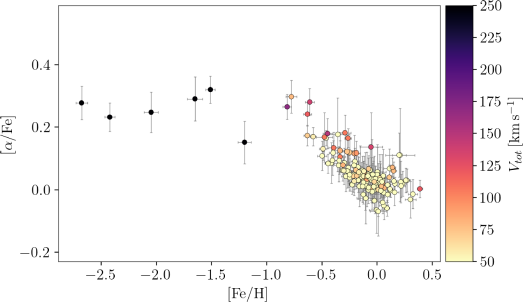}
 \includegraphics[width=0.45\textwidth,clip]{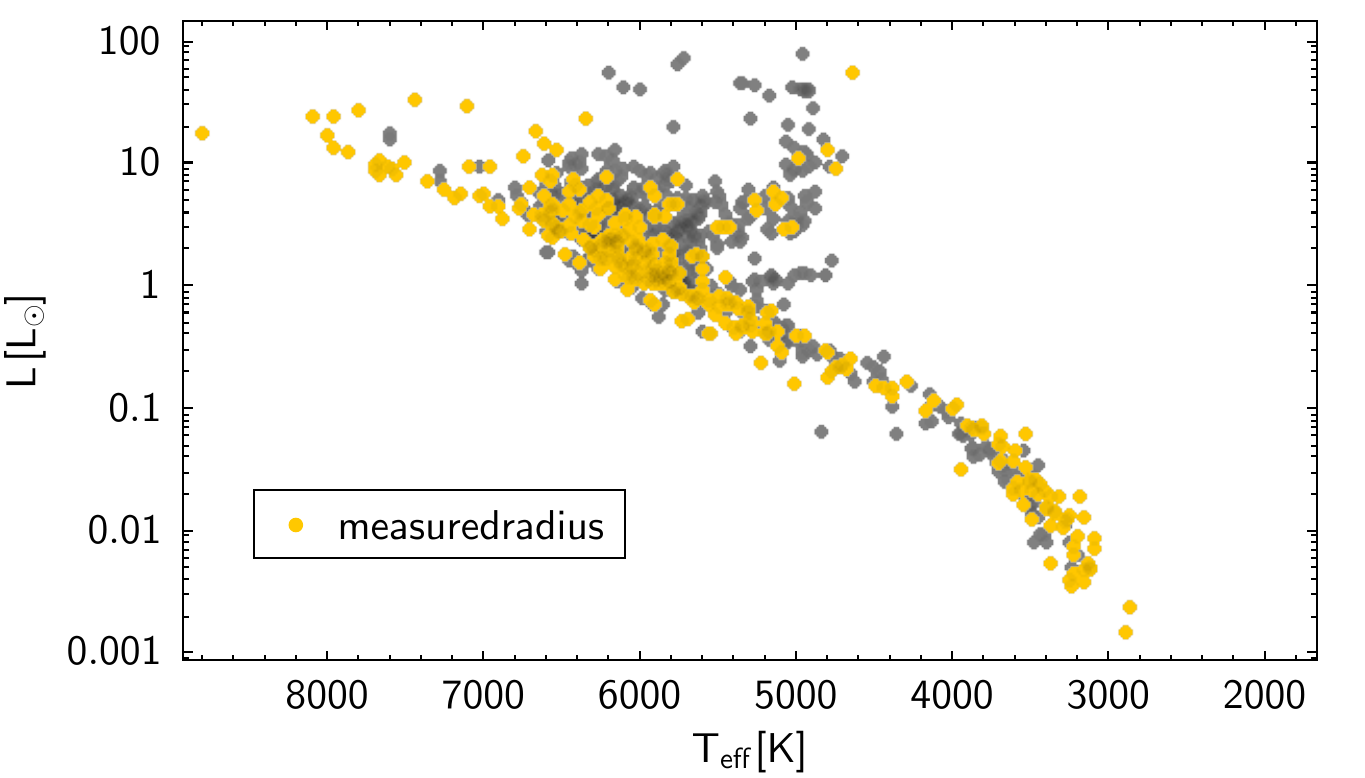}
%% Note the ABSENCE of the extension .pdf  !
  \caption{\textbf{Left:} Variation of [$\alpha$/Fe] with [Fe/H] for the Gaia Benchmark Stars. Credits: \citet{2026A&A...705A.167C}.
  \textbf{Right:} FGKM main sequence and subgiant stars in the PLATO Benchmark stars catalogue.}
  \label{creevey:fig2}
\end{figure}

The Gaia FGK Benchmark Stars collaboration \cite{Jofre2014,Heiter2015} is particularly active in this topic.  The most recent version of their catalogue is version 3 \citep[v3,][]{2024A&A...682A.145S} which exploits existing angular diameters and performs a homogenous analysis of the bolometric flux, \fbol, by collecting photometry from a large set of public catalogues.   The new v3 is a compilation of over 200 benchmark stars with fundamental properties of radius, \teff, luminosity, and surface gravity \logg, with  165 stars having a \teff\ accuracy of 2\%  and \logg\ less than 0.1 dex.   
Using the v3 parameters, \citet{2026A&A...705A.167C} produced a corresponding catalogue of $\alpha$ and iron-peak abundances with typical precisions of about 0.05 dex.  This is illustrated in Fig.~\ref{creevey:fig2} left panel where the resulting [$\alpha$/Fe] is plotted as a function of [Fe/H] for the sample, colour-coded by their total velocity.  As expected high velocity stars correspond to the $\alpha$-rich, metal-poor older populations.

Similar efforts in benchmark stars but extending to M stars and extending to the parameters of mass, radius, and age is also ongoing in the context of the Plato Benchmark Stars workpackage, as part of the calibration and validation efforts within the PLATO consortium \citep{Kaushik2026,Merle2026,2026arXiv260404042Z}.  Fig.~\ref{creevey:fig2} right panel shows the distribution of these stars in the HR diagram, which focusses mainly on main sequence and subgiant stars and extends to the low temperature regime.

%presents fundamental properties of 200 stars by exploiting existing high quality observations of angular diameters from the literature.  They collected photometry from many surveys and performed a homogenous analysis of the spectral energy distribution of the full sample to derive the stellar bolometric flux \fbol. 

\subsection{Gaia}\label{creevey:sec:gaia}
%%-------------------------

The European Space Agency’s Gaia mission is revolutionizing the way we conduct astronomy and our understanding of the Milky Way, with the mission’s data contributing to more than 2,000 scientific papers each year. The Gaia catalog has become a fundamental resource for stellar photometric, astrometric, and kinematic parameters, providing homogeneous measurements for hundreds of millions of stars. Data Release 3 (DR3), published in June 2022, marked a major step forward in this effort. The French astronomical community has been particularly active not only in the production and validation of the catalogue, in particular through its Service National d'Observations\footnote{\url{https://www.insu.cnrs.fr/fr/les-services-nationaux-dobservation}} \citep{DR3-DPACP-159,DR3-DPACP-155,DR3-DPACP-157,DR3-DPACP-186} providing radial velocities, astrophysical parameters and chemical abundances, among others, but also in its scientific exploitation. For example,  through an analysis of a {\it golden sample} from Gaia DR3, \citet{DR3-DPACP-123} has provided new insights into the low-mass end of the main sequence. These stars are intrinsically faint and therefore more difficult to detect, despite accounting for more than 75\% of the stellar population by number. As illustrated in Figure~\ref{creevey:fig4} left panel, the distribution of stellar radius  as a function of \teff\ reveals an apparent inversion close to the stellar -- sub-stellar boundary.  More broadly, Gaia DR3 has provided fundamental stellar properties for hundreds of millions of stars, opening an unprecedented window onto the stellar populations of our Galaxy.

\begin{figure}
 \centering
 \includegraphics[width=1\textwidth,clip]{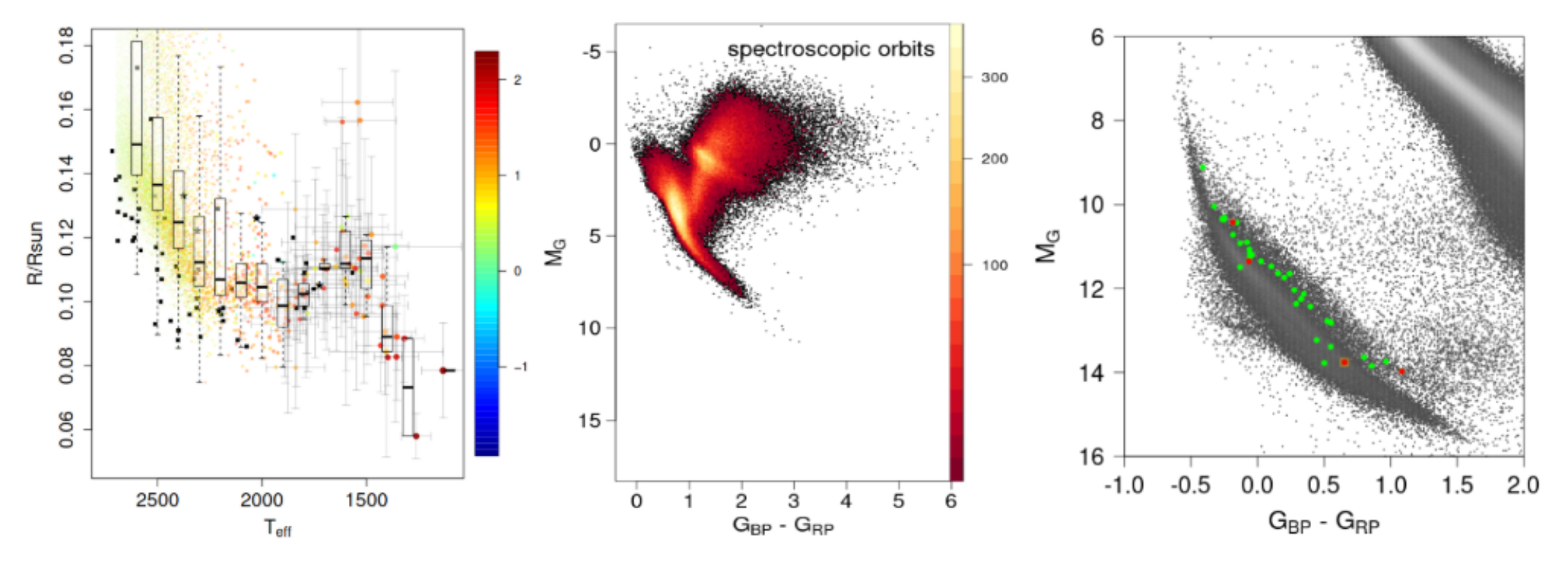}
%% Note the ABSENCE of the extension .pdf  !
  \caption{Examples of contributions of the French community to Gaia and provision of fundamental properties of stars 
  \textbf{Left:} Variation of radius with \teff\ in the low-mass stellar regime.  Credits: \citet{DR3-DPACP-123}.
  \textbf{Center:} Distribution of binaries in Gaia DR3 in spectroscopic orbits.  
  \textbf{Right:} Distribution of binaries in Gaia DR3 with white dwarf companions. Credits (center and right): \citet{DR3-DPACP-100}.
}
  \label{creevey:fig3}
\end{figure}

Of particular interest is the contribution of Gaia to the study of stellar multiplicity \citep{DR3-DPACP-163,DR3-DPACP-100} through astrometric, spectroscopic, and photometric observations.  For example, there are several tens of thousands of binaries detected through spectroscopy, and  Fig.~\ref{creevey:fig3} middle panel illustrates for good quality parallaxes ($<20$\% precision) the distribution of these stars in the colour-magnitude diagram (CMD).
If the binary orbit is completely resolved, for example through a combination of astrometry and spectroscopy, or through photometric eclipses and spectroscopy, one can derive model-independent masses.   This is particularly powerful, because masses are one of the key drivers of stellar evolution, and by constraining this property observationally for two or more components of the same age and initial composition, a very stringent test on the stellar models is possible.  
Table~3 of \citet{DR3-DPACP-100} summarises the systems within Gaia DR3, known at the time of the release, for which either the primary, secondary, or both masses are constrained.  While nearly 200\,000 systems have constraints on the secondary mass (by assuming the primary),  about 80 systems have both masses entirely resolved, and thus provide excellent testbeds for stellar structure, evolution and ages.  
Other ongoing efforts in this respect is presented in \cite{Halbwachs2026}, see also \citet{2018MNRAS.474..731K,2020MNRAS.496.1355H} who exploit the Observatoire Haute Provence, among other observatories to better characterise binaries with the aim to derive their masses.  
Among the binaries in Gaia DR3 with mass constraints are systems with very compact objects and Fig.~\ref{creevey:fig3} right panel illustrates the white dwarfs in such systems in the CMD.  The discovery of black holes with Gaia \citep[][and references therein]{2024A&A...686L...2G}  is one of the  major successes.

\section{Focus on advances in modelling}\label{creevey:sec:model}
\begin{figure}[ht!]
 \centering
 \includegraphics[width=0.98\textwidth,clip]{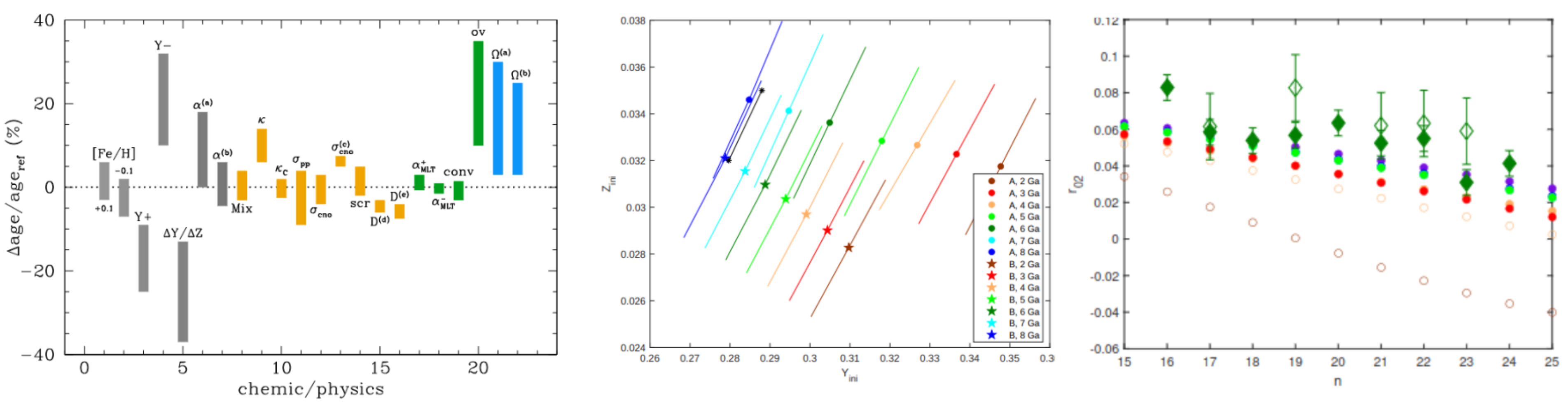}
  \caption{{\bf Left:} Impact of the choice of physics in models on stellar age.  Credits: \citet{2014EAS....65...99L}. 
  {\bf Center:} Combinations of $Z_i,Y_i$ that reproduce the observations of $\alpha$ Cen A and B.  
  {\bf Left:} Comparison of seismic data (open) with model frequencies for different choices of physical model inputs.  
  Credits (center and left): \citet{2026A&A...707A.353T}. }
  \label{creevey:fig4}
\end{figure}

%%---------------------------------
As illustrated in Fig.~\ref{creevey:fig4} left panel, the impact of chemistry, microscopic and macroscopic processes in stellar models can change the age of a star by up to 40\% of its main sequence lifetime \citep[Fig.~43 of][]{2014EAS....65...99L}.  Some of the largest contributors to these uncertainties are the poor knowledge of the initial chemical composition of a star ($Y$, $\alpha$, [Fe/H], where $Y$ is the  initial mass fraction of helium).   In order to test the impact of all of these ingredients, one must use benchmark stars with strong observational constraints, such as asteroseismology, interferometry, abundances, ...   
The $\alpha$ Cen system is one of these benchmark systems, having in addition to high quality observations, dynamical masses, and asteroseismology \citep{2007A&A...470..295B,2017A&A...597A.137K}.

\cite{2026A&A...707A.353T} conducted a focussed test on the $\alpha$ Cen system, where they aimed to derive an age of the system by exploiting the fact that both components should have a unique initial composition ($Z_i,Y_i$), where here $Z$ refers to all elements heavier than helium.  
%Instead of using [Fe/H] as metallicity, the chemical composition in stellar structure models is described by the mass fraction of hydrogren $X$, helium $Y$ and all other metals collectively described by $Z$, where $X+Y+Z=1$, and the typical mixture of the elements (Ca, Fe, N, ...) is assumed to be similar to that of the Sun.   
In their study they explored the combinations of the initial ($Z_i,Y_i$,age) which reproduced the observational constraints.  These are illustrated in the middle panel of Fig.~\ref{creevey:fig4} by the coloured lines representing a combination of $Z_i,Y_i$ at a particular age (brown = 2 Ga, red = 3 Ga, ...).  The top lines denote A while the lower B.  As one can see, as the age increases (going left towards the blue), the lines from both A and B come closer and closer together, with the best overlap denoted by the black line.   The age that they inferred from this analysis is 7.8 Ga, for $Z_i=0.032-0.035$ and $Y_i=0.276-0.288$, for a specific solar mixture and masses.  

The authors then repeated the analysis for several recently debated solar mixtures \citep{1993PhST...47..133G, 2009ARA&A..47..481A} and obtained several combinations of ($Z_i,Y_i$,age) for each mixture.  With these best models, they then compared asteroseismic data with the predicted theoretical frequencies coming from the best combination models.  The comparison between the observed (green with error bars) and theoretical seismic data is shown in the right panel of Fig.~\ref{creevey:fig4}.  The best agreement with the seismic data (purple) favoured the older higher metal composition of $Z/X = 0.245$ over the newer highly debated value of 0.0181, and this preferred model for $\alpha$ Cen A presented no convective core.

These analyses were conducted using the CESAM2K stellar structure and evolution code \citep{1997A&AS..124..597M,2008Ap&SS.316...61M}.  This code has been continuously developed \citep{2013A&A...549A..74M} and is now known as CESAM2k20 \citep{2025A&A...704A..79M} and is the official code for the production of the stellar models to be used in the PLATO pipelines (Deal et al. in prep.).  The new PLATO data should additionally provide more constraints from asteroseismology which will help to refine elements of stellar evolution that remain poorly constrained.  

\begin{figure}[ht!]
 \centering
 \includegraphics[width=0.98\textwidth,clip]{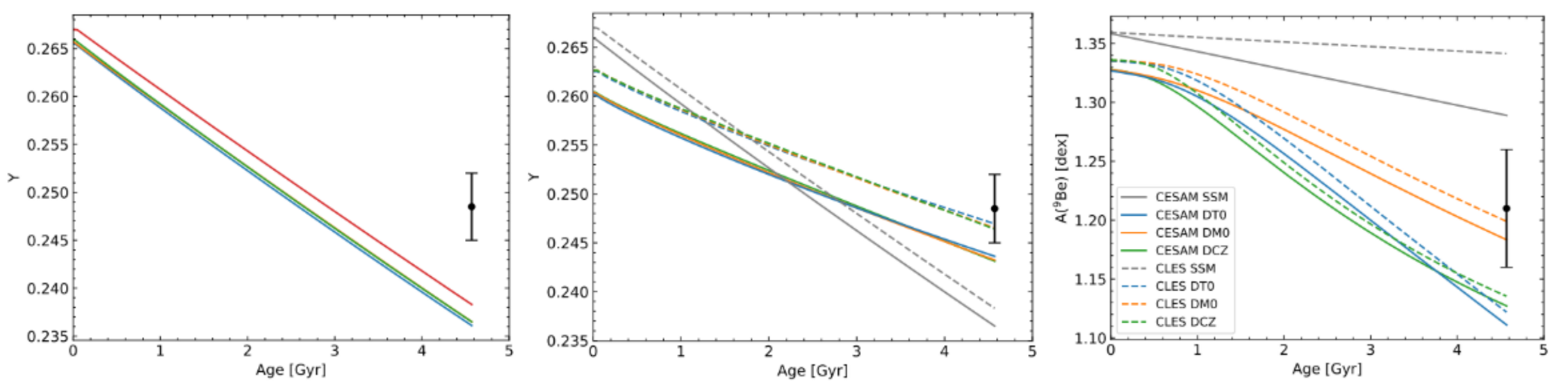}
  \caption{Impact of input physics in stellar models on the observed chemical abundances.  {\bf Left and center:} Evolution of helium mass fraction with different suppositions in the models. {\bf Right:} Evolution of A($^9$Be).  Credits: \citet{2025SoPh..300...96D}.}
  \label{creevey:fig5}
\end{figure}

Exploiting CESAM2K20 along with the CLES stellar evolution code \citep{2008Ap&SS.316..149S}, \citet{2025SoPh..300...96D} investigated the addition of extra mixing processes in models in order to find a solution to the current issue in reproducing the  light elements of the Sun.    Both codes treat atomic diffusion in slightly different ways, and they tested different implementations of a so-called {\it turbulent mixing} that helps to avoid the depletion of helium ($Y$) and heavy elements.  In Fig~\ref{creevey:fig5} left panel, the various implementations of atomic diffusion with both codes is illustrated by following the evolution of the $Y$ with time.  The observed helium abundance today is illustrated by the point with an error bar.  It is clear that with the classical implementation, the models fail to reproduce the observed $Y$ while this issue is not seen with $Z$ (not shown here).  The middle panel of Fig~\ref{creevey:fig5} shows again the evolution of $Y$ but this time by including the various implementations of the ad-hoc mixing (see label of right figure for details).   Here we can see that some of the models are now capable of reproducing the observed $Y$.  However, when the theoretical abundance of $^9$Be is studied (right panel), we can see that only the "DMO" (reference depth of mixing related to mass) implementation in both the CESAM2K20 and CLES codes are capable of reproducing the observed abundance. 
In this paper they further explored other macroscopic mixing processes by also exploiting helioseismic data.

These theoretical experiments are extremely important if we want to push our understanding of the details of stellar processes in stars.  In order to this we need to have benchmark stars with enough quality and accuracy to provide these stringent constraints on the models.   This improvement of models feeds directly back into other topics by providing the optimal theoretical tools to other communities.  In particular, a more accurate knowledge of the ages of astrophysical systems will serve to better understand formation and evolution scenarios.

%The following examples illustrate the required style in the main text:
% \cite{Einstein26} showed that $v_h = 1926 \, \kms$ was not a prime number, as also found previously \citep{Laurel24}. On the other hand, \citet{1945RvMP...17..120E} followed \citet{Kafka24} and found a better approximation to the distance, yielding $d_o = 22 \, \kpc$ \citep[see also][and references therein]{Bohr26,Curie91,deGaulle96}. 
%For frequently cited papers an abbreviated form of citation is recommended, e.g. Paper~I, Paper~II.

%The format for references is the one adopted by A\&A. To set the reference list in the proper format, we encourage you to use \BibTeX~ and the natbib package instead of the standard \verb=\thebibliography= environment.

%\subsection{Figures}
%%-------------------

%Please use files in the PDF format, as shown in Fig.~\ref{author1:fig1}. Figures in eps format can be converted to pdf with the command epstopdf (used as ``epstopdf file.eps''). Two figures can be joined together as shown in Fig.~\ref{author1:fig2}. In this case, as illustrated in Fig.~\ref{author1:fig2},  the caption must follow this format (e.g. boldface fonts for the Left and Right items): \textbf{Left:} text of the caption for the left panel. \textbf{Right:} text of the caption for the figure on the right hand side. 

%%
%% Example of two figures side by side
%%

%%
%% Example of two figures side by side
%%

\section{Prospects and Conclusions}
%%--------------------
In this talk I highlighted the important role that stellar physics plays in our quest for understanding astrophysical processes in a variety of topics, with some concrete examples focussing on the understanding of exoplanets, stars themselves, and the Milky Way.   While interfaces imply a two-way relation, in this work I focussed primarily on the central role that stellar physics plays for other scientific questions.  The main idea presented concerns the importance of deriving fundamental properties of stars in order to provide observational constraints on theoretical processes.  
Several recent examples of both the observational and theoretical efforts were presented.  From the observational point of view, we highlighted the important role of interferometry for providing fundamental properties, and the Gaia mission for providing complementary observations and identification and characterisation of multiple systems.  Future missions, such as PLATO, HWO, and the ELT will help to complete these observational constraints. 
On the theoretical side, many other important works exist in the literature such as the influence of magnetism and rotation, and 3D modelling of stellar atmospheres e.g. \citet{2026exoe.confE..28C,2026A&A...711A.128P, 2026A&A...709A.224D}, that were not addressed here but are of equal importance towards our understanding of processes in stars, and which feed back to other communities.

% Optional acknowledgements
% -------------------------
\begin{acknowledgements}
Thank you to the organizers of the SF2A 2026.
\end{acknowledgements}

\bibliographystyle{aa}  % A&A bibliography style file (aa.bst)
\bibliography{Creevey_S00} % your references in file: Yourfile.bib

\end{document}